\RequirePackage{fix-cm}
\documentclass[smallextended]{svjour3}       
\smartqed  
\usepackage[T1]{fontenc} 

\usepackage{graphicx}
\usepackage{caption}
\usepackage{subcaption}
\usepackage{dcolumn}
\usepackage{bm}
\usepackage{amssymb,amsmath}
\usepackage{amscd,amsfonts}
\usepackage{sansmath}
\usepackage{cite}
\usepackage{fullpage}
\usepackage{float}
\usepackage{lineno}
\usepackage{mathrsfs}
\usepackage{hyperref}
\usepackage{url}
\usepackage{color}
\usepackage{xcolor}
\usepackage{epsfig}
\usepackage{epstopdf}
\usepackage{appendix}
\usepackage{booktabs}
\usepackage{rotating}
\usepackage{multicol}
\usepackage{multirow}
\usepackage{changepage}

\DeclareGraphicsExtensions{.eps,.pdf,.png,.jpg,.jpeg}

\begin{document}

\title{{On the dark matter as a geometric effect in $\mathnormal{f}(\mathcal{R})$ gravity		
}}


\author{Muhammad Usman
}


\institute{Muhammad Usman \at
              School of Natural Sciences (SNS),\\
              National University of Sciences and Technology (NUST),\\
              Sector H-12, Islamabad 44000, Pakistan \\
              \email{muhammad\_usman\_sharif@yahoo.com} \\
              \email{muhammadusman@sns.nust.edu.pk}           
}

\date{Received: 20 July 2015 / Accepted: 15 August 2016 in General Relativity and Gravitation}
\maketitle
\begin{abstract}
A mysterious type of matter is supposed to exist, because the observed rotational velocity curves of particle moving around the galactic center and the expected rotational velocity curves do not match. This type of matter is called dark matter. There are also a number of proposals in the modified gravity which are alternatives to the dark matter. In this contrast, 
in $2008$, Christian G. B$\ddot{\text{o}}$hmer, Tiberiu Harko and Francisco S.N. Lobo presented an interesting idea in \cite{Böhmer2008386} where they showed that a $\mathnormal{f}(\mathcal{R})$ gravity model could actually explain dark matter to be a geometric effect only. They solved the gravitational field equations in vacuum using generic $\mathnormal{f}(\mathcal{R})$ gravity model for constant velocity regions (i.e. dark matter regions around the galaxy). They found that the resulting modifications in the Einstein Hilbert Lagrangian is of the form $\mathcal{R}^{1+m}$, where $m=V_{tg}^2/c^2$; $V_{tg}$ being the tangential velocity of the test particle moving around the galaxy in the dark matter regions and $c$ being the speed of light. From observations it is known that $m\approx\mathcal{O}(10^{-6})$ \cite{Böhmer2008386,Salucci11062007,Persic01071996,Borriello11052001}.
In this article, we perform two things (1) We show that the form of $\mathnormal{f}(\mathcal{R})$ they claimed is not correct. In doing the calculations, we found that when the radial component of the metric for constant velocity regions is a constant then the exact solutions for $\mathnormal{f}(\mathcal{R})$ obtained is of the form of $\mathcal{R}^{1-\alpha}$ which corresponds to a negative correction rather than positive claimed by the authors of \cite{Böhmer2008386}, where $\alpha$ is the function of $m$.
(2) We also show that we can not have an analytic solution of $f(\mathcal{R})$ for all values of tangential velocity including the observed value of tangential velocity $200-300$Km/s \cite{Salucci11062007,Persic01071996,Borriello11052001} if the radial coefficient of the metric which describes the dark matter regions is \emph{not a constant}. 
Thus, we have to rely on the numerical solutions to get an approximate model for dark matter in $\mathnormal{f}(\mathcal{R})$ gravity.
\end{abstract}

\section{Introduction}\label{intro}
%
Newtonian Physics suggests that the tangential velocity of a test particle around the galaxy should be inversely proportional to its radial distance from the center of the galaxy (i.e. $V_{tg}^2=\text{G}M/r$), but for the spiral galaxies the tangential velocity of a particle around galactic center seems to remain approximately constant as we move away from the galactic center \cite{vera1,vera2}. Standard Newtonian Physics does not have any answer to this observation.

Einstein's gravity which is linear in Ricci scalar $\mathcal{R}$ suggests that either there is some invisible matter present around the galaxy (thus we should put that in the right hand side (matter part) of the Einstein field equations) or Einstein's gravity is not a viable gravity model around the spiral galaxies which could explain the observed rotational velocity curves.

To explain this unexpected observations, there are many proposals. A few of them are: modified Newtonian dynamics \cite{milgrom}, modified gravity \cite{Böhmer2008386,mannheim}, weakly interacting massive particles (WIMP) \cite{jungman} etc.

In $\mathnormal{f}(\mathcal{R})$ modified gravity models, the Einstein Hilbert Lagrangian is replaced by an arbitrary function of the Ricci scalar $\mathcal{R}$. Some times this arbitrary function is computed for some observed phenomenon and some times for a given (inspired) $\mathnormal{f}(\mathcal{R})$ function the dynamics of objects or system of objects or the Universe is analyzed.

In $2008$, Christian G. B$\ddot{\text{o}}$hmer, Tiberiu Harko and Francisco S.N. Lobo 
solved the gravitational field equations in vacuum (i.e. $T_{\mu\nu}=0$) using arbitrary function of Ricci scalar $\mathcal{R}$ 
for dark matter regions around the galaxy. They tried to solve the field equations for an exact solution and found that $\mathnormal{f}(\mathcal{R})=\mathcal{R}^{1+m}$ with $m=V_{tg}^2/c^2$ can explain the unorthodox behavior of the spiral galaxies rotation velocity curves as to be the geometric effects only.

In this article, we review their work and present some new results. After going through the same procedure they adopted, we find that the form of $\mathnormal{f}(\mathcal{R})$ they claim to get is not correct.
We show that in the same scheme two different sets of parameters are obtained which then give two different forms of $f(\mathcal{R})$ function that can explain the constant tangential velocity curve but both forms indicate that rather than having a positive correction of the form $\mathcal{R}^{1+\alpha}$, a negative correction is more favored, of the form $\mathcal{R}^{1-\alpha}$. We also obtained the numerical results of the field equations in \cite{Usman} to check our findings of the current paper and found the results given here to be consistent with the numerical results.


The article is organized as follows. In the next section, we present the field equations for $\mathnormal{f}(\mathcal{R})$ gravity in vacuum. In section 3, we try to find out the exact solution of the field equations derived in section 2. Conclusions and discussions on the results/findings of section 3 are in section 4.
\section{$\mathnormal{f}(\mathcal{R})$ gravity field equations in vacuum}\label{fRfieldequations}
The metric which describes the static and spherically symmetry in the constant velocity regions is
\begin{equation}\label{eq:metric}
{\mathrm{d}S}^2=-\mathnormal{e}^{\nu(r)}{\mathrm{d}t}^2+\mathnormal{e}^{\lambda(r)}{\mathrm{d}r}^2+r^2{\mathrm{d}\theta}^2+r^2{\sin\theta}^2{\mathrm{d}\phi}^2
\end{equation}
where $\nu(r)=2m ~ \mathnormal{ln}(r/r_0)$, here $r_0$ is the constant of integration obtained in deriving the coefficient $\nu(r)$ of the metric.

The $f(\mathcal{R})$ modified gravity action is written as
\begin{equation}\label{eq:action}
S=\int \sqrt{-g} \mathnormal{f}(\mathcal{R})\mathrm{d}^4x~,
\end{equation}
where $\mathnormal{f}(\mathcal{R})$ is an arbitrary analytic function of Ricci scalar $\mathcal{R}$. The variation of the above action with respect to the metric $g_{\mu\nu}$ gives the following field equations
\begin{equation}\label{eq:fieldequations}
F(\mathcal{R})\mathcal{R}_{\mu\nu}-\dfrac{1}{2}\mathnormal{f}(\mathcal{R})\mathnormal{g}_{\mu\nu}-\left( \nabla_\mu\nabla_\nu-\mathnormal{g}_{\mu\nu}\square \right)F(\mathcal{R})=0~,
\end{equation}
where $F(\mathcal{R})=\mathrm{d}\mathnormal{f}/\mathrm{d}\mathcal{R}$. The contraction of the above equation gives
\begin{equation}\label{eq:fieldequationscontraction}
F(\mathcal{R})\mathcal{R}-2\mathnormal{f}(\mathcal{R})+3\square F(\mathcal{R})=0~.
\end{equation}
Upon using eq. (\ref{eq:fieldequationscontraction}) in eq. (\ref{eq:fieldequations}) we get modified field equations as \cite{PhysRevD.74.064022},
\begin{equation}\label{eq:modifiedfieldequations}
F(\mathcal{R})\mathcal{R}_{\mu\nu}-\nabla_\mu\nabla_\nu F=\dfrac{1}{4}\mathnormal{g}_{\mu\nu}\left( F\mathcal{R}-\square F \right)~,
\end{equation}
also differentiation of eq. (\ref{eq:fieldequationscontraction}) with respect to `r' gives,
\begin{equation}
\mathcal{R}F^{'}-\mathcal{R}^{'}F+3\left(\square F\right)^{'}=0.
\end{equation}
Where {\large{$'$}} represents derivative with respect to `r'. If we use modified field equations to find out the solution then the solution must also satisfy eq. (\ref{eq:fieldequationscontraction}) to be a solution of original field equations given by eq. (\ref{eq:fieldequations}). From eq. (\ref{eq:modifiedfieldequations}), we see that $A_\mu\equiv (F\mathcal{R}_{\mu\mu}-\nabla_\mu\nabla_\mu F)/\mathnormal{g}_{\mu\mu}$ is independent of the index $\mu$, thus $A_\mu-A_\nu=0$. From the relation $A_\mu-A_\nu=0$ we can write three equations ($A_0-A_1=0,~ A_1-A_2=0\text{ and }A_0-A_2=0$). 
The equations $A_0-A_1=0,~ A_1-A_2=0$, the $rr$ component of the field equations (\ref{eq:fieldequations}) and eq. (\ref{eq:fieldequationscontraction}) can be written as \cite{refId0-1,refId0-2} \footnote{Detailed derivation of the result $A_\mu-A_\nu=0$ and subsequent equations has been done in the appendix \ref{appendix:A}},
\begin{equation}\label{eq:1st}
F^{''}-\dfrac{1}{2}(\nu^{'}+\lambda^{'})F^{'}+\dfrac{1}{r}(\nu^{'}+\lambda^{'})F=0~,
\end{equation}
\begin{equation}\begin{split}\label{eq:2nd}
\nu^{''}+{\nu^{'}}^2-\dfrac{1}{2}\left(\nu^{'}+\dfrac{2}{r}\right)
\left(\nu^{'}+\lambda^{'}\right)-\dfrac{2}{r^2}+2\dfrac{\mathnormal{e}^{\lambda}}{r^2}
=2\dfrac{F^{''}}{F}-\left(\lambda^{'}+\dfrac{2}{r}\right)\dfrac{F^{'}}{F}~,
\end{split}\end{equation}
\begin{equation}\label{eq:3rd}
\begin{array}{rcl}
\mathnormal{f}=F\mathnormal{e}^{-\lambda}\left[\nu^{''}-\dfrac{1}{2}\nu^{'}\left(\nu^{'}+\lambda^{'}\right)- 2\dfrac{\lambda^{'}}{r}\right. 
+\left.\left(\nu^{'}+\dfrac{4}{r}\right)\dfrac{F^{'}}{F} \right]~,
\end{array}
\end{equation}
\begin{equation}\label{eq:4th}
\mathcal{R}=2\dfrac{f}{F}-3\mathnormal{e}^{-\lambda}\left( \dfrac{F^{''}}{F}+\left( \dfrac{1}{2}\left(\nu^{'}-\lambda^{'}\right)+\dfrac{2}{r} \right)\dfrac{F^{'}}{F} \right)~,
\end{equation}
respectively. Introducing  $\eta=\ln(r/r^{*})$, Eq. (\ref{eq:1st})-(\ref{eq:4th}) now become
\begin{equation}\label{eq:5th-eta}
	\dfrac{\mathrm{d}^2F}{\mathrm{d}\eta^2}-\left[1+\dfrac{1}{2}\left(\dfrac{\mathrm{d}\nu}{\mathrm{d}\eta}+\dfrac{\mathrm{d}\lambda}{\mathrm{d}\eta}\right)\right]\dfrac{\mathrm{d}F}{\mathrm{d}\eta}+\left(\dfrac{\mathrm{d}\nu}{\mathrm{d}\eta}+\dfrac{\mathrm{d}\lambda}{\mathrm{d}\eta}\right)F=0~,
\end{equation}
\begin{equation}\label{eq:6th-eta}
	\dfrac{\mathrm{d}^2\nu}{\mathrm{d}\eta^2}-\dfrac{\mathrm{d}\nu}{\mathrm{d}\eta}+\left(\dfrac{\mathrm{d}\nu}{\mathrm{d}\eta}\right)^2-\dfrac{1}{2}\left(\dfrac{\mathrm{d}\nu}{\mathrm{d}\eta}+2\right)\left(\dfrac{\mathrm{d}\nu}{\mathrm{d}\eta}+\dfrac{\mathrm{d}\lambda}{\mathrm{d}\eta}\right)-2\left(1-e^\lambda\right)=2\dfrac{1}{F}\dfrac{\mathrm{d}^2F}{\mathrm{d}\eta^2}-\left(\dfrac{\mathrm{d}\lambda}{\mathrm{d}\eta}+4\right)\dfrac{1}{F}\dfrac{\mathrm{d}F}{\mathrm{d}\eta}~,
\end{equation}
\vspace{0.18cm}
\begin{equation}\label{eq:7th-eta}
	f=\dfrac{F\mathnormal{e}^{-\lambda-2\eta}}{{r^{*}}^2}\left[\dfrac{\mathrm{d}^2\nu}{\mathrm{d}\eta^2}-\dfrac{\mathrm{d}\nu}{\mathrm{d}\eta}-\dfrac{1}{2}\left(\dfrac{\mathrm{d}\nu}{\mathrm{d}\eta}+\dfrac{\mathrm{d}\lambda}{\mathrm{d}\eta}\right)\dfrac{\mathrm{d}\nu}{\mathrm{d}\eta}\right.
	-2\dfrac{\mathrm{d}\lambda}{\mathrm{d}\eta}+\left.\left(\dfrac{\mathrm{d}\nu}{\mathrm{d}\eta}+4\right)\dfrac{1}{F}\dfrac{\mathrm{d}F}{\mathrm{d}\eta}\right],
\end{equation}
\vspace{0.18cm}
\begin{equation}\label{eq:8th-eta}
	\mathcal{R}=2\dfrac{f}{F}-\dfrac{3\mathnormal{e}^{-\lambda-2\eta}}{{r^{*}}^2}\left[ \dfrac{1}{F}\dfrac{\mathrm{d}^2F}{\mathrm{d}\eta^2}\right.-\dfrac{1}{F}\dfrac{\mathrm{d}F}{\mathrm{d}\eta}
	+\left( \dfrac{1}{2}\left(\dfrac{\mathrm{d}\nu}{\mathrm{d}\eta}-\dfrac{\mathrm{d}\lambda}{\mathrm{d}\eta}\right)+2 \right)\left.\dfrac{1}{F}\dfrac{\mathrm{d}F}{\mathrm{d}\eta}\right],
\end{equation}
Now introducing
\begin{equation}\label{eq:u}
1/F\left(\mathrm{d}F/\mathrm{d}\eta\right)=u~.
\end{equation}
to reduce the order in $F$, eqs. (\ref{eq:5th-eta}) and (\ref{eq:6th-eta}) become
\begin{equation}\label{eq:eq7}
\dfrac{\mathrm{d}u}{\mathrm{d}\eta}+u^2-\left(1+\dfrac{1}{2}\left(\dfrac{\mathrm{d}\nu}{\mathrm{d}\eta}+\dfrac{\mathrm{d}\lambda}{\mathrm{d}\eta}\right)\right)u+\dfrac{\mathrm{d}\nu}{\mathrm{d}\eta}+\dfrac{\mathrm{d}\lambda}{\mathrm{d}\eta}=0~,
\end{equation}
\begin{align}\begin{split}\label{eq:eq8}
\dfrac{\mathrm{d}^2\nu}{{\mathrm{d}\eta}^2}-\dfrac{\mathrm{d}\nu}{\mathrm{d}\eta}+\left(\dfrac{\mathrm{d}\nu}{\mathrm{d}\eta}\right)^2&-\dfrac{1}{2}\left(\dfrac{\mathrm{d}\nu}{\mathrm{d}\eta}+2\right)\left(\dfrac{\mathrm{d}\nu}{\mathrm{d}\eta}+\dfrac{\mathrm{d}\lambda}{\mathrm{d}\eta}\right) 
-2\left(1-\mathnormal{e}^{\lambda}\right)=2\left(\dfrac{\mathrm{d}u}{\mathrm{d}\eta}+u^2\right)-\left(\dfrac{\mathrm{d}\lambda}{\mathrm{d}\eta}+4\right)u~,
\end{split}\end{align}
using ${\mathrm{d}u}/{\mathrm{d}\eta}+u^2$ from eq. (\ref{eq:eq7}) in eq. (\ref{eq:eq8}) we get
\begin{align}\begin{split}\label{eq:eq12->11}
\dfrac{\mathrm{d}^2\nu}{{\mathrm{d}\eta}^2}-\dfrac{\mathrm{d}\nu}{\mathrm{d}\eta}+\left(\dfrac{\mathrm{d}\nu}{\mathrm{d}\eta}\right)^2&-\left(1+\dfrac{1}{2}\dfrac{\mathrm{d}\nu}{\mathrm{d}\eta}\right)\left(\dfrac{\mathrm{d}\nu}{\mathrm{d}\eta}+\dfrac{\mathrm{d}\lambda}{\mathrm{d}\eta}\right) \\ & 
-2\left(1-\mathnormal{e}^{\lambda}\right)+2\left(1-\dfrac{1}{2}\dfrac{\mathrm{d}\nu}{\mathrm{d}\eta}\right)u+2\left(\dfrac{\mathrm{d}\nu}{\mathrm{d}\eta}+\dfrac{\mathrm{d}\lambda}{\mathrm{d}\eta}\right)=0~.
\end{split}\end{align}
Any two equations of eqs. (\ref{eq:eq7})-(\ref{eq:eq12->11}) provide the solution of $\lambda$ and $u$ ($\nu$ is known).
\section{
	Dark matter as a geometric effect}\label{darkmatter}
To obtain the geometric interpretation of the constant velocity regions, we solve eq. (\ref{eq:eq7}) and eq. (\ref{eq:eq12->11}) for $\lambda$ and $u$. With $\nu=2m(\eta-\eta_0)$, eq. (\ref{eq:eq7}) and eq. (\ref{eq:eq12->11}) can now be written as
\begin{equation}\label{eq:DMeq1}
\dfrac{\mathrm{d}u}{\mathrm{d}\eta}+u^2-\left(1+m+\dfrac{1}{2}\dfrac{\mathrm{d}\lambda}{\mathrm{d}\eta}\right)u+2m+\dfrac{\mathrm{d}\lambda}{\mathrm{d}\eta}=0~,
\end{equation}
\begin{equation}\label{eq:DMeq2}
-2m+4m^2+\left(1-m\right)\left(2m+\dfrac{\mathrm{d}\lambda}{\mathrm{d}\eta}\right)+2\left(1-m\right)u-2\left(1-\mathnormal{e}^{\lambda}\right)=0~.
\end{equation}
As discussed in \cite{Böhmer2008386}, the tangential velocity of a test particle in stable circular orbit around the galactic center is about $200-300\text{Km/s}$ \cite{Salucci11062007,Persic01071996,Borriello11052001} thus $m\approx\mathcal{O}(10^{-6})$, this allows us to neglect terms containing $m^i$ where $i\geq2$ in eqs. (\ref{eq:DMeq1}), (\ref{eq:DMeq2}) and in all subsequent equations. A particular case considered by B$\ddot{\text{o}}$hmer et al. in \cite{Böhmer2008386} to solve the above equation is $\lambda=$constant.
\subsection{When $\lambda$ is a constant}
The eq. (\ref{eq:DMeq1}) and (\ref{eq:DMeq2}) now becomes
\begin{equation}\label{eq:DMeqlambdaconstant1}
	\dfrac{\mathrm{d}u}{\mathrm{d}\eta}+u^2-\left(1+m\right)u+2m=0~,
\end{equation}
\begin{equation}\label{eq:DMeqlambdaconstant2}
	m^2+\left(1-m\right)u-\left(1-\mathnormal{e}^{\lambda}\right)=0~.
\end{equation}
When $\lambda=\text{constant}$, eq. (\ref{eq:DMeqlambdaconstant2}) implies that $u$ is also a constant. From eq. (\ref{eq:DMeqlambdaconstant1}) we obtain
$$u=\{ 2m, 1-m \}~,$$
using the above values of $u$ in eq. (\ref{eq:DMeqlambdaconstant2}) we obtain
$$e^\lambda=\{ 1-2m, 2m \}$$ respectively.
\subsubsection{When $u=2m,\text{\space}e^\lambda=1-2m$}
With the above parameters we get from eq. (\ref{eq:u})
\begin{equation}\label{eq:fr1}
F(r)=F_0 ~ r^{2m}~,
\end{equation}
where $F_0=F_c/{r^{*}}^{2m}$, $F_c$ is the constant of integration.
Now from eqs. (\ref{eq:3rd}) and (\ref{eq:4th})
\begin{equation}\label{fR1f}
f(r)=6m ~ F_0 ~ r^{2(m-1)}
\end{equation}
\begin{equation}\label{fR1R}
\mathcal{R}(r)=\dfrac{2m}{r^2}
\end{equation}
Using $r$ from eq. (\ref{fR1R}) into eq. (\ref{fR1f}) we obtain $f$ in the parametric form as
\begin{equation}\label{fR1}
f(\mathcal{R})=3(2m)^{m} ~ F_0 ~ \mathcal{R}^{1-m}
\end{equation}
The above equation shows that we should expect $f(\mathcal{R})$ to be less than $\mathcal{R}$ for constant tangential velocity regions around the galaxy.
\subsubsection{When $u=1-m,\text{\space}e^\lambda=2m$}
With the above parameters we get from eq. (\ref{eq:u})
\begin{equation}\label{eq:fr2}
F(r)=F_0 r^{1-m}~,
\end{equation}
where $F_0=F_c/{r^{*}}^{(1-m)}$, $F_c$ is the constant of integration.
Now from eqs. (\ref{eq:3rd}) and (\ref{eq:4th})
\begin{equation}\label{fR2f}
f(r)=3 ~ F_0 ~ r^{-(1+m)}
\end{equation}
\begin{equation}\label{fR2R}
\mathcal{R}=\dfrac{3}{r^2}
\end{equation}
Using $r$ from eq. (\ref{fR2R}) into eq. (\ref{fR2f}) we obtain $f$ in the parametric form as
\begin{equation}\label{fR2}
f(\mathcal{R})={3}^{(1-m)/2} ~ F_0 ~ \mathcal{R}^{1-(1-m)/2}
\end{equation}
The above equation again tells us that we should expect a negative correction to the Einstein's general relativity for constant tangential velocity regions around the galaxy.
\subsection{When $\lambda$ is not a constant}
Using $u$ and $\mathrm{d}u/\mathrm{d}\eta$ from eq. (\ref{eq:DMeq2}) in eq. (\ref{eq:DMeq1}) we obtain
\begin{align}\begin{split}\label{eq:DMeq3}
\dfrac{1}{2}\dfrac{\mathrm{d}^2\lambda}{{\mathrm{d}\eta}^2}&-\dfrac{1}{1-m}\left(\dfrac{1}{2}\mathnormal{e}^\lambda-m(1-m)\right)\dfrac{\mathrm{d}\lambda}{\mathrm{d}\eta}-\dfrac{1}{2}\left(\dfrac{\mathrm{d}\lambda}{\mathrm{d}\eta}\right)^2 \\ &-\dfrac{(1-\mathnormal{e}^\lambda)^2}{(1-m)^2}-\dfrac{\mathnormal{e}^\lambda(1+m^2)-1}{(1-m)^2}-2m=0~,
\end{split}\end{align}
This is a second order differential equation in $\lambda$. After obtaining $\lambda$ from this equation one can use either eq. (\ref{eq:DMeq1}) or eq. (\ref{eq:DMeq2}) to obtain $u$. The obtained $\lambda$ and $u$ must satisfy the third equation (which has not been used in getting $\lambda$ and $u$).
Eq. (\ref{eq:DMeq3}) now up to $\mathcal{O}(m^2)$ can be written as
\begin{align}\begin{split}\label{eq:DMeq4}
\dfrac{1}{2}\dfrac{\mathrm{d}^2\lambda}{{\mathrm{d}\eta}^2}&-\left(\dfrac{1}{2}\left(1+m\right)\mathnormal{e}^\lambda-m\right)\dfrac{\mathrm{d}\lambda}{\mathrm{d}\eta}-\dfrac{1}{2}\left(\dfrac{\mathrm{d}\lambda}{\mathrm{d}\eta}\right)^2 \\ &-\left(1+2m\right){\left(1-\mathnormal{e}^\lambda\right)^2}-\left(1+2m\right)\left(\mathnormal{e}^\lambda-1\right)-2m=0~.
\end{split}\end{align}

\begin{figure}[H]
	\centerline{
		\includegraphics[width=12.5cm]{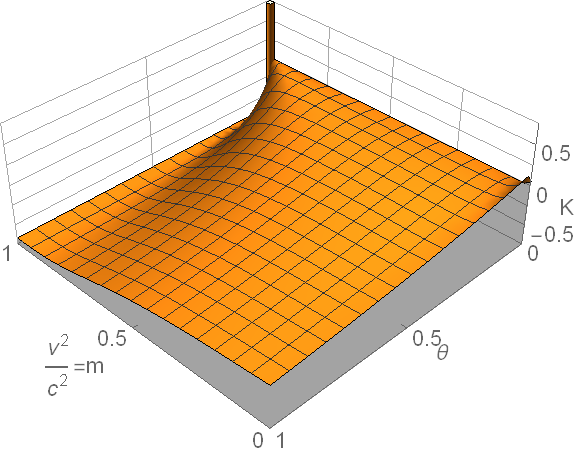}}
	\caption[The plot of $K$]{$K$ as a function of $\theta$ and $m$.}
	\label{fig:K}
\end{figure}

Introducing new variable as $\mathrm{d}\lambda/\mathrm{d}\eta=\rho$ $\implies$ $\mathrm{d}/\mathrm{d}\eta=\rho~\mathrm{d}/\mathrm{d}\lambda$. This allows us to write eq. (\ref{eq:DMeq3}) as
\begin{align}\begin{split}\label{eq:DMnewvarible}
\dfrac{\rho}{2}~\dfrac{\mathrm{d}\rho}{\mathrm{d}\lambda}+\dfrac{1}{1-m}\left(\dfrac{3}{2}\right.\mathnormal{e}^\lambda&-m(1-m){\bigg)}\rho-\dfrac{1}{2}\rho^2 \\ &-\dfrac{(1-\mathnormal{e}^\lambda)^2}{(1-m)^2}-\dfrac{\mathnormal{e}^\lambda(1+m^2)-1}{(1-m)^2}+2m=0~.
\end{split}\end{align}
Now, using the transformation $\rho=1/\omega$ and $e^{\lambda}=\theta$, eq. (\ref{eq:DMnewvarible}) after simplification can be written as
\begin{align}\begin{split}\label{eq:DMtransformed}
\dfrac{\mathrm{d}\omega}{\mathrm{d}\theta}=\left[ \dfrac{m}{\theta}\right.&-\dfrac{2(1-\theta)^2}{\theta(1-m)^2}-\left.\dfrac{2(\theta(1+m^2)-1)}{\theta(1-m)^2} \right]\omega^3
+\left[ 3-\dfrac{2m(1-m)}{\theta} \right]\omega^2-\dfrac{1}{\theta}\omega~.
\end{split}\end{align}
This is the nonlinear first order Abel differential of first kind of the form $$\dfrac{\mathrm{d}\omega}{\mathrm{d}\theta}=p(\theta)\omega^3+q(\theta)\omega^2+r(\theta)\omega+s(\theta)~,$$
with
$$p(\theta)=\dfrac{m}{\theta}-\dfrac{2(1-\theta)^2}{\theta(1-m)^2}-\dfrac{2(\theta(1+m^2)-1)}{\theta(1-m)^2}~,$$
$$q(\theta)=3-\dfrac{2m(1-m)}{\theta} \text{\space , \quad} r(\theta)=-\dfrac{1}{\theta} ~\text{\space , \quad} s(\theta)=0.$$
If we choose that the particular solution $y_p$ of the above equation satisfies
\begin{equation}\begin{split}\label{transformationcondition}
3\left[ \dfrac{m}{\theta}\right.&-\dfrac{2(1-\theta)^2}{\theta(1-m)^2}-\left.\dfrac{2(\theta(1+m^2)-1)}{\theta(1-m)^2} \right] y_p^2 
+2\left[ 3-\dfrac{2m(1-m)}{\theta} \right] y_p-\dfrac{1}{\theta}=0~,
\end{split}\end{equation}
then eq. (\ref{eq:DMtransformed}) becomes Abel differential equation of second kind \cite{Mak200291}, of the form
\begin{align}\begin{split}\label{eq:DMtransformedAbel2nd}
\dfrac{\mathrm{d}\omega}{\mathrm{d}\theta}&=\left[ \dfrac{m}{\theta}-\dfrac{2(1-\theta)^2}{\theta(1-m)^2}-\dfrac{2(\theta(1+m^2)-1)}{\theta(1-m)^2} \right]\omega^3 \\&~~~
+\left[3\left(\dfrac{m}{\theta}-\dfrac{2(1-\theta)^2}{\theta(1-m)^2}-\dfrac{2(\theta(1+m^2)-1)}{\theta(1-m)^2}\right)y_p\right.
+\left.\left(3-\dfrac{2m(1-m)}{\theta}\right) \right]\omega^2 ~.
\end{split}\end{align}
In the case when,
\begin{equation}\label{solutioncondition}
\dfrac{\mathrm{d}}{\mathrm{d}\theta}\left(\dfrac{p(\theta)}{\sqrt{q(\theta)^2-3p(\theta)r(\theta)}}\right)=K\sqrt{q(\theta)^2-3p(\theta)r(\theta)}~,
\end{equation}
where \emph{$K$ is an arbitrary constant} then the general solution of eq. (\ref{eq:DMtransformed}) is given by \cite{Mak200291},
\begin{equation}
\begin{split}
\omega(\theta)=&\pm\dfrac{\sqrt{q(\theta)^2-3p(\theta)r(\theta)}}{p(\theta)}V
+\left(\dfrac{-q(\theta)\pm\sqrt{q(\theta)^2-3p(\theta)r(\theta)}}{3p(\theta)}\right),
\end{split}
\end{equation}
where $V$ satisfies the separable differential equation
\begin{equation}
\dfrac{\mathrm{d}V}{\mathrm{d}\theta}=\left(V^3+V^2+K~V\right)\left(\dfrac{q(\theta)^2-3p(\theta)r(\theta)}{p(\theta)}\right).
\end{equation}
Thus, in order to obtain the analytic solution, we must satisfy the condition given by eq. (\ref{solutioncondition}). The plot of 
$$K=\dfrac{1}{\sqrt{q(\theta)^2-3p(\theta)r(\theta)}}~{\dfrac{\mathrm{d}}{\mathrm{d}\theta}\left(\dfrac{p(\theta)}{\sqrt{q(\theta)^2-3p(\theta)r(\theta)}}\right)}$$ 
is given in fig. (\ref{fig:K}). The plot allows us to understand when (i.e. for what $m$ and $\theta$) could we get an analytic solution to the eq. (\ref{eq:DMtransformed}).
From the fig. (\ref{fig:K}) we observe that When $\theta=0$ we have eq. (\ref{solutioncondition}) satisfied for approximately all values of $m$. This demands an arbitrary negative large value of $\lambda(r)$. When we have $m=1$ eq. (\ref{solutioncondition}) is obeyed for also approximately all values of $\theta$ except $-\infty$. For all other values of $m$ if we take the domain of $\theta$ to be $0\leq\theta\leq1$ \footnote{i.e. all possible values, ($0,\infty$) is being mapped onto ($0,1$) with the mapping function $e^{-1/x}$} then we observe the violation of eq. (\ref{solutioncondition}) (i.e. $K$ is not a constant). Since the observed $m$ is $\mathcal{O}(10^{-6})$, from the fig (\ref{fig:K}) we see that obtaining analytic solution is not possible around the desired value. The plot also shows that we cannot obtain an analytic solution for all values of $m$.
\section{Conclusions and discussion}\label{conclustions}
In concluding the work, we adopted the same procedure as done in \cite{Böhmer2008386} and found that the $\mathnormal{f}(\mathcal{R})$ gravity model given by B$\ddot{\text{o}}$hmer et al in \cite{Böhmer2008386} is not consistent with our results when radial coefficient of the metric describing dark matter region is a constant. We obtained two different forms of $f(\mathcal{R})$ function given by eq. (\ref{fR1}) and (\ref{fR2}) that can explain the constant tangential velocity curve but both forms indicate that the rather than having a positive correction of the form $\mathcal{R}^{1+\alpha}$, a negative correction is more likely to be the real picture of the form $\mathcal{R}^{1-\alpha}$, where $\alpha$ is the function of $m$.

We also extended the same procedure when $\lambda$ is not a constant and found that we can not have an analytic solution to the $f(\mathcal{R})$ gravity model for all values of $m$ to explain the constant rotational velocity curves.

We also obtained the numerical results of the field equations in \cite{Usman} to check our findings of the current paper and found consistency with the numerical results.
\section{Acknowledgment}
This work is supported by {\textit{National University of Sciences and Technology (NUST), Sector H-12 Islamabad 44000, Pakistan}}.
\bibliographystyle{elsarticle-num}
\bibliography{Muhammad-Usman.bib}

\appendix
\renewcommand\appendixtocname{Appendix}
\section{: Derivation of the equations}\label{appendix:A}
As discussed, from eq. (\ref{eq:modifiedfieldequations}), we see that $A_\mu\equiv (F\mathcal{R}_{\mu\mu}-\nabla_\mu\nabla_\mu F)/\mathnormal{g}_{\mu\mu}$ is independent of the index $\mu$, thus $A_\mu-A_\nu=0$. The proof is as follows: eq. (\ref{eq:modifiedfieldequations}) can be written as
\begin{equation}\label{appendix:eq:modifiedfieldequations1}
	F(\mathcal{R})\mathcal{R}_{\mu\nu}-\nabla_\mu\nabla_\nu F(\mathcal{R})=\dfrac{1}{4}g_{\mu\nu}\left( F(\mathcal{R})g^{\alpha\beta}\mathcal{R}_{\alpha\beta}-g^{\alpha\beta}\nabla_\alpha\nabla_\beta F(\mathcal{R}) \right)
\end{equation}
\begin{equation}\label{appendix:eq:modifiedfieldequations2}
F(\mathcal{R})\mathcal{R}_{\mu\nu}-\nabla_\mu\nabla_\nu F(\mathcal{R})=\dfrac{1}{4}g_{\mu\nu}g^{\alpha\beta}\Big( F(\mathcal{R})\mathcal{R}_{\alpha\beta}-\nabla_\alpha\nabla_\beta F(\mathcal{R}) \Big)
\end{equation}
Since the metric considered is static and spherically symmetric (\ref{eq:metric}) with diagonal entries thus $\mu=\nu$ and $\alpha=\beta$. Equation (\ref{appendix:eq:modifiedfieldequations2}) now becomes
\begin{equation}\label{appendix:eq:modifiedfieldequations3}
F(\mathcal{R})\mathcal{R}_{\mu\mu}-\nabla_\mu\nabla_\mu F(\mathcal{R})=\dfrac{1}{4}g_{\mu\mu}g^{\alpha\alpha}\Big( F(\mathcal{R})\mathcal{R}_{\alpha\alpha}-\nabla_\alpha\nabla_\alpha F(\mathcal{R}) \Big)
\end{equation}
\begin{equation}\label{appendix:eq:modifiedfieldequations4}
\dfrac{1}{g_{\mu\mu}}\Big(F(\mathcal{R})\mathcal{R}_{\mu\mu}-\nabla_\mu\nabla_\mu F(\mathcal{R})\Big)=\dfrac{1}{4}g^{\alpha\alpha}\Big( F(\mathcal{R})\mathcal{R}_{\alpha\alpha}-\nabla_\alpha\nabla_\alpha F(\mathcal{R}) \Big)
\end{equation}
\begin{equation}\label{appendix:eq:modifiedfieldequations5}
\dfrac{1}{g_{\mu\mu}}\Big(F(\mathcal{R})\mathcal{R}_{\mu\mu}-\nabla_\mu\nabla_\mu F(\mathcal{R})\Big)=\dfrac{1}{4 g_{\alpha\alpha}}g_{\alpha\alpha}g^{\alpha\alpha}\Big( F(\mathcal{R})\mathcal{R}_{\alpha\alpha}-\nabla_\alpha\nabla_\alpha F(\mathcal{R}) \Big)
\end{equation}
Since, $g^{\alpha\alpha}g_{\alpha\alpha}=4$. Thus, above equation now becomes
\begin{equation}\label{appendix:eq:modifiedfieldequations6}
\dfrac{1}{g_{\mu\mu}}\Big(F(\mathcal{R})\mathcal{R}_{\mu\mu}-\nabla_\mu\nabla_\mu F(\mathcal{R})\Big)=\dfrac{1}{g_{\alpha\alpha}}\Big( F(\mathcal{R})\mathcal{R}_{\alpha\alpha}-\nabla_\alpha\nabla_\alpha F(\mathcal{R}) \Big)
\end{equation}
Since $\alpha$ is free index, we can replace it by $\nu$. Thus,
\begin{equation}\label{appendix:eq:modifiedfieldequations7}
\dfrac{1}{g_{\mu\mu}}\Big(F(\mathcal{R})\mathcal{R}_{\mu\mu}-\nabla_\mu\nabla_\mu F(\mathcal{R})\Big)=\dfrac{1}{g_{\nu\nu}}\Big( F(\mathcal{R})\mathcal{R}_{\nu\nu}-\nabla_\nu\nabla_\nu F(\mathcal{R}) \Big)
\end{equation}
\begin{equation}\label{appendix:eq:modifiedfieldequations8}
A_\mu=A_\nu \text{\qquad\qquad\qquad for all $\mu$ and $\nu$,}
\end{equation}
where $A_\mu=\dfrac{1}{g_{\mu\mu}}\Big(F(\mathcal{R})\mathcal{R}_{\mu\mu}-\nabla_\mu\nabla_\mu F(\mathcal{R})\Big)$ and $A_\nu=\dfrac{1}{g_{\nu\nu}}\Big(F(\mathcal{R})\mathcal{R}_{\nu\nu}-\nabla_\nu\nabla_\nu F(\mathcal{R})\Big)$. Here, $\nabla_i\nabla_j=\partial_i\partial_j+\Gamma^{\sigma}_{ij}$.

From the above relation $A_\mu-A_\nu=0$ we can write three equations ($A_0-A_1=0,~ A_1-A_2=0\text{ and }A_0-A_2=0$; `$0$' means `$t$', `$1$' means `$r$', `$2$' means `$\theta$' and `$3$' means `$\phi$'). \footnote{Since the field equations also give three equation we can use any three equations from all six equations.}

The equation $A_0-A_1=0$ can be written as
\begin{equation}\label{appendix:eq:modifiedfieldequations_A0-A1-1}
\dfrac{1}{g_{00}}\Big(F(\mathcal{R})\mathcal{R}_{00}-\nabla_0\nabla_0 F(\mathcal{R})\Big)=\dfrac{1}{g_{11}}\Big(F(\mathcal{R})\mathcal{R}_{11}-\nabla_1\nabla_1 F(\mathcal{R})\Big)
\end{equation}
As the metric in consideration is static, eq. (\ref{appendix:eq:modifiedfieldequations_A0-A1-1}) now becomes
\begin{equation}\label{appendix:eq:modifiedfieldequations_A0-A1-2}
\dfrac{F,_{11}}{g_{11}}+\left( \dfrac{\mathcal{R}_{00}}{g_{00}}-\dfrac{\mathcal{R_{11}}}{g_{11}} \right)F-\left( -\dfrac{\Gamma^{1}_{00}}{g_{00}}+\dfrac{\Gamma^{1}_{11}}{g_{11}} \right)F,_{1}=0
\end{equation}
This leads to eq. (\ref{eq:1st}). The equation $A_1-A_2=0$ can be written as
\begin{equation}\label{appendix:eq:modifiedfieldequations_A1-A2-1}
\dfrac{1}{g_{11}}\Big(F(\mathcal{R})\mathcal{R}_{11}-\nabla_1\nabla_1 F(\mathcal{R})\Big)=\dfrac{1}{g_{22}}\Big(F(\mathcal{R})\mathcal{R}_{22}-\nabla_2\nabla_2 F(\mathcal{R})\Big)
\end{equation}
This leads to eq. (\ref{eq:2nd}). The $rr$ component of the field equations (\ref{eq:fieldequations}) leads to eq. (\ref{eq:3rd}). Equation (\ref{eq:fieldequationscontraction}) leads to eq. (\ref{eq:4th})

\section{: Comparison of equations and results with \cite{Böhmer2008386}}\label{appendix:B}
In this section, we perform a comparison of our equations and results with the B$\ddot{\text{o}}$hmer et. al. \cite{Böhmer2008386} \footnote{Only which had error in \cite{Böhmer2008386}.}.
\begin{table}[H]
\renewcommand{\arraystretch}{3}
	\begin{adjustwidth}{-1cm}{}
	\centering
	\label{tab:table1}
	\begin{tabular}{|l|c|l|c|}
	\hline 
	\multicolumn{2}{|c|}{Correct equations} & \multicolumn{2}{|c|}{B$\ddot{\text{o}}$hmer et. al. \cite{Böhmer2008386}} \\ 
	\hline \hline 
	\multirow{2}{*}{Equation (\ref{eq:1st})} & \multirow{2}{*}{$F^{''}-\dfrac{1}{2}(\nu^{'}+\lambda^{'})F^{'}+\dfrac{1}{r}(\nu^{'}+\lambda^{'})F=0~$} & \multirow{2}{*}{Equation (18)} & \multirow{2}{*}{$F^{''}-\dfrac{1}{2}(\nu^{'}+\lambda^{'})F^{'}\textbf{-}\dfrac{1}{r}(\nu^{'}+\lambda^{'})F=0~$} \\ & & &  \\ \hline
	\multirow{2}{*}{Equation (\ref{eq:2nd})}& $\nu^{''}+{\nu^{'}}^2-\dfrac{1}{2}\left(\nu^{'}+\dfrac{2}{r}\right)\left(\nu^{'}+\lambda^{'}\right)-\dfrac{2}{r^2}$ & \multirow{2}{*}{Equation (19)} & $\nu^{''}+{\nu^{'}}^2-\dfrac{1}{2}\left(\nu^{'}+\dfrac{2}{r}\right)\left(\nu^{'}+\lambda^{'}\right)-\dfrac{2}{r^2}$ \\ & $~~~~~~~~~~~+2\dfrac{\mathnormal{e}^{\lambda}}{r^2}=2\dfrac{F^{''}}{F}-\left(\lambda^{'}+\dfrac{2}{r}\right)\dfrac{F^{'}}{F}$ & & $~~~~~~~~~~~+2\dfrac{\mathnormal{e}^{\lambda}}{r^2}=\textbf{-}2\dfrac{F^{''}}{F}\mathbf{+}\left(\lambda^{'}+\dfrac{2}{r}\right)\dfrac{F^{'}}{F}$ \\
	\hline 
	\multirow{2}{*}{Equation (\ref{eq:5th-eta})} & $\dfrac{\mathrm{d}^2F}{\mathrm{d}\eta^2}-\left[1+\dfrac{1}{2}\left(\dfrac{\mathrm{d}\nu}{\mathrm{d}\eta}+\dfrac{\mathrm{d}\lambda}{\mathrm{d}\eta}\right)\right]\dfrac{\mathrm{d}F}{\mathrm{d}\eta}$ & \multirow{2}{*}{Equation (23)} & $\dfrac{\mathrm{d}^2F}{\mathrm{d}\eta^2}-\left[1+\dfrac{1}{2}\left(\dfrac{\mathrm{d}\nu}{\mathrm{d}\eta}+\dfrac{\mathrm{d}\lambda}{\mathrm{d}\eta}\right)\right]\dfrac{\mathrm{d}F}{\mathrm{d}\eta}$ \\ 
	& $~~~~~~~~~~+\left(\dfrac{\mathrm{d}\nu}{\mathrm{d}\eta}+\dfrac{\mathrm{d}\lambda}{\mathrm{d}\eta}\right)F=0$ & & $~~~~~~~~~~-\left(\dfrac{\mathrm{d}\nu}{\mathrm{d}\eta}+\dfrac{\mathrm{d}\lambda}{\mathrm{d}\eta}\right)F=0$ \\
	\hline
	\multirow{2}{*}{Equation (\ref{eq:6th-eta})} & $\dfrac{\mathrm{d}^2\nu}{\mathrm{d}\eta^2}-\dfrac{\mathrm{d}\nu}{\mathrm{d}\eta}+\left(\dfrac{\mathrm{d}\nu}{\mathrm{d}\eta}\right)^2-\dfrac{1}{2}\left(\dfrac{\mathrm{d}\nu}{\mathrm{d}\eta}+2\right)\left(\dfrac{\mathrm{d}\nu}{\mathrm{d}\eta}+\dfrac{\mathrm{d}\lambda}{\mathrm{d}\eta}\right)$ & \multirow{2}{*}{Equation (24)} & $\dfrac{\mathrm{d}^2\nu}{\mathrm{d}\eta^2}-\dfrac{\mathrm{d}\nu}{\mathrm{d}\eta}+\left(\dfrac{\mathrm{d}\nu}{\mathrm{d}\eta}\right)^2-\dfrac{1}{2}\left(\dfrac{\mathrm{d}\nu}{\mathrm{d}\eta}+2\right)\left(\dfrac{\mathrm{d}\nu}{\mathrm{d}\eta}+\dfrac{\mathrm{d}\lambda}{\mathrm{d}\eta}\right)$ \\
	& $-2\left(1-e^\lambda\right)=2\dfrac{1}{F}\dfrac{\mathrm{d}^2F}{\mathrm{d}\eta^2}-\left(\dfrac{\mathrm{d}\lambda}{\mathrm{d}\eta}+4\right)\dfrac{1}{F}\dfrac{\mathrm{d}F}{\mathrm{d}\eta}$ & & $\textbf{+}2\left(1-e^\lambda\right)=\textbf{-}2\dfrac{1}{F}\dfrac{\mathrm{d}^2F}{\mathrm{d}\eta^2}\textbf{+}\left(\dfrac{\mathrm{d}\lambda}{\mathrm{d}\eta}+4\right)\dfrac{1}{F}\dfrac{\mathrm{d}F}{\mathrm{d}\eta}$ \\ \hline
	\end{tabular}
	\caption{Comparison of equations}
\end{adjustwidth}
\end{table}
\subsection{Comparison of results}
B$\ddot{\text{o}}$hmer et. al. in \cite{Böhmer2008386} gave only \emph{one} $f(\mathcal{R})$ gravity function, $f(\mathcal{R})=f_0 \mathcal{R}^{(1+m)}$, as being the replacement of the dark matter when in fact there should have been two function as eq. (18) in \cite{Böhmer2008386} is second order differential equation.
On the other hand, we found two different $f(\mathcal{R})$ gravity functions given by eq.  (\ref{fR1}) and (\ref{fR2}).

\end{document}